
\newif\ifOmitFigures   


\ifOmitFigures\message{FIGURES WILL BE OMITTED}\else\message{FIGURES WILL
BE INCLUDED}\fi \ifOmitFigures\def\input#1#2#3#4{{}}\def\epsfverbosetrue{}
\def\special#1{\vskip 5 truecm}\def\epsfbox#1{\vskip 5 truecm}\fi
\newdimen\epsfxsize\newdimen\epsfysize




%
 \let\miguu=\footnote
 \def\footnote#1#2{{$\,$\parindent=9pt\baselineskip=13pt%
 \miguu{#1}{#2\vskip -7truept}}}
%
%


\def\implies{\Rightarrow}
\def\=>{\Rightarrow}
\def\==>{\Longrightarrow}
 
 \def\dal{\displaystyle{{\hbox to 0pt{$\sqcup$\hss}}\sqcap}}
 
%
\def\lto{\mathop
        {\hbox{${\lower3.8pt\hbox{$<$}}\atop{\raise0.2pt\hbox{$\sim$}}$}}}
\def\gto{\mathop
        {\hbox{${\lower3.8pt\hbox{$>$}}\atop{\raise0.2pt\hbox{$\sim$}}$}}}
%
 



\def\less{\backslash}		


\def\hat{\widehat}		




\def				
  \Complexes
   {{\rm C}\llap{\vrule height6.3pt width1pt depth-.4pt\phantom t}}




\def\interior #1 {  \buildrel\circ\over  #1}     



\input epsf
\epsfverbosetrue

\raggedbottom	 		
\magnification=\magstep1	

\hsize=6 true in
 \hoffset=0.27 true in
\vsize=8.5 true in
 \voffset=0.28 true in

\def\~{\widetilde}  
\def\H{{\cal H}}
\def\future{{\rm future}}
\def\cut{\hfil\break}
\def\hat{\widehat}
\def\Nobreak{\par\nobreak}

\def\singlespace{\baselineskip=12pt}  
\def\sesquispace{\baselineskip=14pt}	

\sesquispace               	


\centerline{\bf Quantum Measure Theory and its Interpretation } 


\singlespace  
\bigskip
 \centerline {\it Rafael D. Sorkin}
 \smallskip
 \centerline {\it Department of Physics, 
                 Syracuse University, 
                 Syracuse, NY 13244-1130} 
 \smallskip
 \centerline {\it and}
 \smallskip 
 \centerline {\it Instituto de Ciencias Nucleares, 
                 UNAM, A. Postal 70-543,
                 D.F. 04510, Mexico}

 \centerline {\it \qquad\qquad internet address: rdsorkin@mailbox.syr.edu}

\singlespace                                
\bigskip
\leftskip=1.5truecm\rightskip=1.5truecm     
\centerline{\bf Abstract}
\medskip                         
\noindent

We propose a realistic, spacetime interpretation of quantum theory in which
reality constitutes a {\it single} history obeying a ``law of motion'' that
makes definite, but incomplete, predictions about its behavior.  We
associate a ``quantum measure'' |S| to the set S of histories, and point
out that |S| fulfills a sum rule generalizing that of classical probability
theory.  We interpret |S| as a ``propensity'', making this precise by
stating a criterion for |S|=0 to imply ``preclusion'' (meaning that the
true history will not lie in S).  The criterion involves triads of
correlated events, and in application to electron-electron scattering, for
example, it yields definite predictions about the electron trajectories
themselves, independently of any measuring devices which might or might not
be present.  (In this way, we can give an objective account of
measurements.)  Two unfinished aspects of the interpretation involve {\it
conditonal} preclusion (which apparently requires a notion of
coarse-graining for its formulation) and the need to ``locate spacetime
regions in advance'' without the aid of a fixed background metric (which
can be achieved in the context of conditional preclusion via a construction
which makes sense both in continuum gravity and in the discrete setting of
causal set theory).

\bigskip
\leftskip=0truecm\rightskip=0truecm         
\sesquispace                                
\bigskip\medskip

\bigskip
\noindent{\bf Three Principles}
\Nobreak

Let me begin by listing three principles and asking whether or not they are
compatible with quantum mechanical practice (as opposed to one or another
interpretation of a particular mathematical version of the quantum
formalism).  There are many reasons for raising such a question, but to my
mind the most important is the need to construct an interpretive framework
for quantum gravity, which I believe we will attain only by holding onto
these principles (cf. [1]).  In the present talk, I will try to
convince you that this belief is viable by sketching an interpretative
framework for quantum mechanics in general, which honors the principles,
but which will still allow us to continue using quantum theory in the
manner to which we have grown accustomed.

The three principles in question are those of {\it Realism}, of the {\it
Spacetime Character} of reality, and of the {\it Single World}.

By realism/objectivity I mean for example that in electron-electron
scattering, the electrons exist and have definite trajectories, and that
consequently a statement of the form ``the electrons never scatter at 90
degrees'' is meaningful in itself, without needing to be reinterpreted as 
shorthand for ``If we set up detectors at 90 degrees they will never
register an event''.

By the spacetime- (or better the ``history-'') character of reality, I mean
for example that it is the 4-metric $^4g_{ab}$ which is real and not the
``wave-function of the universe'' $\psi$, or for that matter, some purely
spatial positive-definite metric $^3g_{jk}$.

By the principle of the single world, I mean for example that in an
electron diffraction experiment, the electron traverses a single definite
slit, and not somehow two slits simultaneously, or one slit in one
``world'' and another slit in some other world.  But ``real'' and
``single'' does not entail ``deterministic'', and I do not mean to deny
that (on current evidence) the world is fundamentally stochastic---so that,
for example, the electron's past trajectory does not determine fully what
its future trajectory will be.

The motivation for these principles comes partly from familiar philosophical
worries and partly from the projected  needs of quantum
gravity.  The so-called Copenhagen Interpretation has no answer to the
question ``Who shall observe the observer?'', no way to give a rational
account of ``wave function collapse'', and more generally no escape from
the vicious circle that we ourselves are made out of atoms and therefore
cannot be more real than they are (cf. [2]).  In the early
universe---one of the main anticipated fields of application of quantum
gravity---these questions assume a much greater practical importance,
because then there were no observers at all, and having to refer to that
era through the indirect medium of present-day observations would complicate
unbearably the already difficult questions of quantum cosmology.  For
quantum gravity more generally, its fundamental diffeomorphism-invariance
means that only global properties of the metric have physical meaning, and
it is hard to see how such properties could be reduced to statements about
objects tied to spacelike hypersurfaces, like the wave-function and
spatial metric of canonical quantum gravity.  Hence the need for a
spacetime character.  As for the ``singleness''of the world, it is hard for
me to imagine what the contrary hypothesis might mean, or how physics can
have any predictive content at all if everything conceivable actually
occurs.\footnote{*}
{Beyond this general remark, I find it hard to comment explicitly on the
``many worlds'' interpretation, because I have never achieved a clear
understanding of what it is intended to mean.  If I had to put my finger on
what seems most obscure to me, I probably would choose to emphasize the
so-called ``pointer basis problem''.}

But do the facts of quantum mechanics actually allow us to hold on to these
three principles?  Perhaps the strongest argument that they must, is merely
an appeal to the obvious truth that we (or at least most of us) {\it
experience} a single world of really existing objects extended in
spacetime.  If this is in fact the nature of our experience, then logically
it ought not to be possible to force us to some other
viewpoint---especially if the character of that other viewpoint is
precisely to elevate our subjective experience above what we naively take
to be objective reality.  However convincing such reasoning might be,
though, it does not yet suggest concretely how a realistic, spacetime
account of quantum mechanics would go.  For that, we must look to the
theory itself.

\bigskip
\noindent{\bf The Sum-over-histories}
\Nobreak

Of the existing formulations of quantum mechanics, the only one which
provides a starting point for constructing a spacetime framework is of
course the sum-over-histories formulation, which in fact is explicitly
called a ``spacetime approach'' in one of the founding papers on the
subject [3].  In that formulation, the central dynamical quantity is
what I will call the ``quantum measure''\footnote{$^\dagger$}
{Not to be confused with the so-called measure-factor in the path integral,
which corresponds to only part of what is here meant by the measure.}
$|S|$ of a subset $S$ of the space $\H$ of all possible histories
$\gamma$ [3][4][5].  Here ``history'' is a
synonym for ``possible reality'', the concrete meaning of which depends on
what one takes to be the basic form of matter one is dealing with: a
collection of particle world-lines or spacetime fields; a 4-dimensional
Lorentzian manifold; a causal set [6]; or whatever.\footnote{*}
{Notice that this usage is in one way much more restrictive than that of
 reference [7], where arbitrary sequences of projection operators
 qualify as ``histories''.  Notice as well, that a change in how one
 identifies the basic ``substance'' will in general change the meaning of
 the principle that one and only one history is realized.  Thus, when in
 the introduction I illustrated this principle with the statement that the
 electron traverses only a single slit, I was assuming that electrons and
 spacetime itself are, if not fundamental, then emergent in such a way that
 the concept of electron trajectory continues to make sense as a property
 of a single history.  By way of contrast, the analogous assumption would
 seem considerably less warranted in the case of photon trajectories, for
 example, which have to be reinterpreted in terms of field configurations
 if one takes the electromagnetic field to be the fundamental reality.}
I will maintain that the task of understanding quantum mechanics in a
manner compatible with the three principles enunciated above will be
accomplished if we succeed in understanding the meaning of the quantum
measure in a way which frees it from any reference to the customary
apparatus of expectation-values and wave-function collapse.

In non-relativistic quantum mechanics or quantum field theory, the measure
$|S|$ of a particular subset $S$ of the space of all histories $\H$
is a nonnegative real number computed by a rather peculiar looking rule
[4] [7] [8], which involves not just individual
``paths'' or ``histories'', but {\it pairs} of them, corresponding to the
familiar fact that probabilities turn out to be quadratic rather than
linear in the basic amplitudes.  Specifically the rule is
$$
   |S| = \sum\limits_{\gamma_1,\gamma_2\in S_T} \Re D(\gamma_1,\gamma_2),
   \eqno (1)
$$
where initial conditions are assumed to be given at some early time (before
any time to which the properties defining $S$ refer), and one has
introduced a {\it truncation time} (or hypersurface) $T$ which is late
enough to be after any time to which the properties defining $S$ refer.
Then $\gamma_1$ and $\gamma_2$ are truncated histories which emerge from
the initial spacetime region with some (joint) amplitude $\alpha$ and
propagate to $T$; and $D(\gamma_1,\gamma_2)$ is either zero or the product
of $\alpha$ with the amplitude of $\gamma_2$ times the complex conjugate of
the amplitude of $\gamma_1$, according as the paths $\gamma_1$ and
$\gamma_2$ either do not or do come together at $T$ (in the field theory
case ``coming together'' means having equal restriction to the hypersurface
$T$).  Pictorially this expression appears as in figure 1.  The
corresponding formula in operator language is $|S|={\rm tr}\; C^* \rho_i C
$ where $\rho_i$ is an initial density operator and $C$ is the operator
which propagates $\rho$ forward when the domain of integration in the
path-integral is restricted to just those $\gamma$ belonging to $S$.
(Notice that the real-part operator $\Re$ can be omitted without changing
the value of the sum (1).) 

 \epsfysize=2.0in
 \centerline{\vbox{\epsfbox{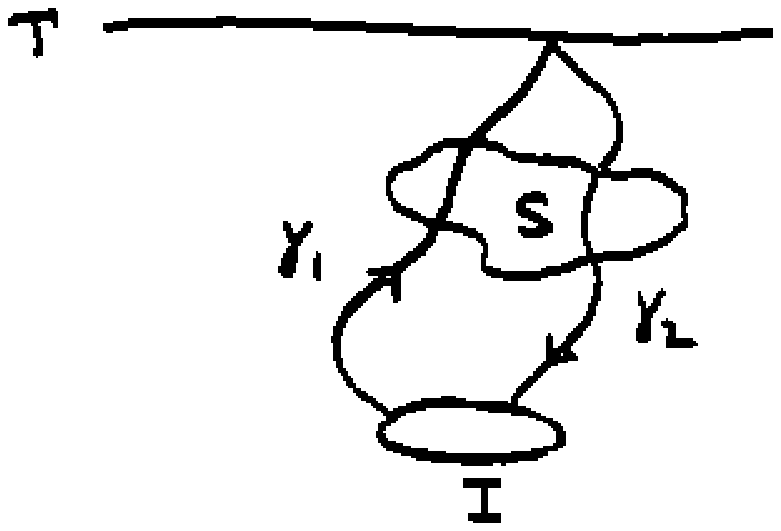}}}
\noindent{\bf Figure 1} Definition of the quantum measure in non-relativistic
                        quantum mechanics 
\vskip .25in

In the special case that the set $S$ is defined by requiring the history to
belong to specified regions of configuration space at specified moments of
time, $|S|$ reduces to the usual quantum mechanical probability that the
corresponding position-measurements will all yield affirmative results, but
this special case can exhaust the physical meaning of the measure only to
the extent that we retreat from a fully spacetime picture to something like
a pre-relativistic one, thereby running head-on into what is often called
the ``problem of time in quantum gravity.''  To avoid such an eventuality,
we must find an interpretation of $|S|$ which frankly adopts a spacetime
standpoint rather than appealing to some notion of position-measurements at
specified times.


To grasp the meaning of quantum mechanics from such a standpoint means
first of all to understand why the measure assumes the peculiar
``quadratic'' form that it does, and second of all --- and more importantly
--- to specify the physical meaning of the measure $|\cdot|$ (in effect its
predictive content) without appealing to the notion of observation or
measurement as an undefined primitive of the theory.  Let us take up these
two questions in turn.

\bigskip
\noindent{\bf Quantum Measure Theory}
\par\nobreak

Seen in the way I am advocating, quantum mechanics is a direct descendent
of the classical theory of stochastic processes [9], and differs
from it only in that a different ``probability calculus'' is involved,
namely that of classical measure theory (a point stressed early on in
[10]).  Classical measure theory also attaches a number $|S| \ge 0$ to
each (measurable) set of histories, but there $|S|$ has an immediate
interpretation as a probability, and accordingly obeys the classical sum
rule (with `$\sqcup$' denoting disjoint union),
$$ 
     I_2(A,B) := |A \sqcup B| - |A| -|B| = 0,         \eqno (2) 
$$ 
which allows the measure to be given a frequency interpretation.

  In the quantum generalization, $|S|$ is still $\ge 0$ but (2) gets
weakened to the following pair of axioms, any solution of which may be
called a {\it quantum measure}.
$$
       |N| = 0 \;  \implies \;  |A \sqcup N| = |A|  \eqno (3)
$$
$$
       I_3(A,B,C) := |A \sqcup B \sqcup C| 
                     - |A \sqcup B| - |A \sqcup C| - |B \sqcup C| 
                    + |A| +|B| +|C| 
                   = 0                                        \eqno (4)
$$
Of these two axioms, only the first seems clearly essential for the type of
interpretation to be given below.  The second on the other hand, makes the
quantum measure what it is, and can be thought of concretely in terms of
the ``3-slit experiment'', or more generally in connection with any process
in which three mutually exclusive alternatives interfere [11].  Indeed,
the 3-slit-experiment can be said to epitomize quantum mechanics by
illustrating the possibility of interference on one hand (i.e. the fact
that $I_2$ does not always vanish), and on the other hand
the fact that
no {\it new} type of interference arises when one passes from two
alternatives to three or more.

    In this sense, quantum mechanics is a rather mild generalization of
classical measure theory, and (4) is only the first of an infinite
hierarchy of successively more general sum-rules,
$$ 
   I_2=0 \; \implies \;  I_3=0 \;  \implies \;  I_4=0 \; \implies \; \cdots, 
$$ 
each formed using the same pattern of even minus odd, and each of which
might serve as the basis for a further generalization of the quantum
formalism.~[11]   An experimental search for 3-alternative interference,
therefore, would provide a ``null test'' of one of the hallmarks of
quantum mechanics, and---were the test to fail---might suggest the need for
generalizing (4) in the direction of one of the higher
sum-rules.\footnote{$^\dagger$}
{These sum-rules appear to have a theoretical relevance entirely
independent of whether there exist physical processes which directly embody
them.  As pointed out by D. Meyer [12], such processes, if they
existed, could be used to solve in polynomial time certain classes of
problems which would take exponentially long using classical or quantum
computations.  Indeed, there seems to be an entire hierarchy of new
computational complexity classes, corresponding to the  sum
rules $I_n=0$ for $n=4,5,6,\ldots$.}


Here, however, I want to emphasize not the possibility of further
generalization, but the extent to which the sum-rule (4)  accounts
for the ``quadratic character'' of the quantum measure as expressed in
(1).  In fact one can prove that, given any set-function $|\cdot|$
for which $I_3(A,B,C)$ vanishes identically, the definition
$$
     I(A,B) := |A \cup B| + |A \cap B| - |A \less B| - |B \less A|
$$
provides a function of pairs of sets which: $(i)$ extends the function
$I_2$ of eq. (2) to overlapping arguments; $(ii)$ is additive
separately in each argument:
$$ 
    I(A \sqcup B, C) = I(A,C) + I(B,C); 
$$ 
and $(iii)$ allows $|\cdot|$ itself to be recovered via the
equality
$$ |A| = {1 \over 2} I(A,A). $$ 
[Thus $I(A,B)$ corresponds to (twice) the real part of the ``decoherence
functional'' $D(A,B)$.]

In this sense, the fact that the quantum measure can be derived from a
``decoherence function'' is explained by the sum-rule (4).  However
(4) by itself does not explain the appearance in quantum theory of
complex numbers---together with the associated unitarity---nor does it
entail the ``Markov'' character of the measure (meaning the fact that
amplitudes evolve locally in time via the Schr\"odinger equation).  On the
other hand realistic quantum systems are in practice always open, and
therefore their evolution is strictly speaking neither unitary nor
Markovian (cf. [13], [14] and the first reference in
[5]).  Thus, the extra generality afforded by the ``quantum
measure theory'' framework is, in practice, already needed for the correct
description of everyday systems, whatever may be its ultimate fate in
connection with quantum gravity.

\bigskip
\noindent{\bf The Meaning of the Measure}
\Nobreak
\smallskip

We saw above that the quantum measure reduces in special cases to a
quantity which one could consistently interpret as the probability of 
an affirmative outcome of a sequence of position measurements.  More generally
one can presumably think of $|S|$ as a kind of ``propensity of
realization'' of the set $S\subseteq\H$.   We might expect,
for example, that the ratio $|S| \, / \, |\H\less{}S|$ 
indicates how much
``more likely'' the actual history is to be found in $S$ than in its
complement.  But since $|\cdot|$ does not in general obey the classical sum
rule (2), it is not a probability in the ordinary sense, and it is not
obvious how to make such a notion of ``more likely''
quantitative in a meaningful and consistent manner.
In pondering this task, it seems appropriate to enquire more closely into
what is meant by probability 
in the classical case.

Even classically, probability is hard to understand because it seems to
govern how a history {\it will} develop, but in retrospect is nowhere
visible in how the history actually {\it has} developed.  (The chance of
rain as of yesterday was only 20\%, but today it is raining.)  In some hard
to define sense, probability refers to that which doesn't exist as well as
to that which does, a characteristic which is reflected in the fact that we
often speak of it in terms of imaginary ensembles of universes.

One way to extract a positive meaning out of probabilistic assertions
without invoking an infinite ensemble of universes is to appeal to the
notion of {\it preclusion}\footnote{*}
{This term is borrowed from [15] and belongs to the
 interpretation of probability which seems to go by the name of ``Cournot's
 Principle''.},
which is meant to express the idea that certain events are so unlikely as
to be ``essentially impossible''.  We may say, for example, that in a trial
of ten thousand tosses of a fair coin, it is ``precluded'' that heads will
come up ten thousand times.  More generally, we can try to substitute the
concept of preclusion for that of probability, and seek the dynamical
content of a theory in its statements of preclusion.  In this way, the
predictions we can make become ``definite'' but
incomplete.\footnote{$^\dagger$}
 {Unfortunately, they also become imprecise to the extent that the criterion
 for preclusion is itself imprecise, which it necessarily is classically,
 and quite possibly also must be quantum mechanically.  More generally, it
 may be that one loses something by reducing probability to preclusion, but
 at least the resulting concept is relatively precise and objective.}
That is, the statement that a certain subset $S \subseteq \H$ is
precluded, means that the {\it actual} history $\gamma$ will not belong to
$S$; it is thus a {\it definite} assertion about $\gamma$, but not one that
determines $\gamma$ in all its details, as a prediction in celestial
mechanics would, for example.

Having decided to interpret probability in terms of preclusion, one still
has the further task of incorporating the fact that what is precluded is
not fixed once and for all, but rather ``changes with time''; or in other
words the fact that the future is {\it conditioned on} the past.  (Thus, it
may be that in 1858 the American Civil War was still not inevitable, but
that peace became precluded when John Brown was hanged---at least this was
an implication of his gallows address).  To do this one needs to be able to
make {\it conditional} statements of the form, ``If the past has such and
such properties then such and such a future possibility is precluded''.
With a classical probability measure, the criterion for conditional
preclusion can be derived from the criterion for absolute preclusion just
by relativizing the measure to the appropriate subspace of $\H$.  With a
quantum measure, things will be more complicated.

So far in this section we have mainly been reviewing how one can interpret
{\it classical} probability-measures in terms of preclusion.  The challenge
now is to find a similar scheme for interpreting the quantum measure
$|\cdot|$, a scheme powerful enough to allow us to make the
predictions on whose success our belief in quantum mechanics is based.
In the remainder of this paper, I will propose a candidate scheme of this
kind.  It is a bit complicated, but in its own way natural, and perhaps
even consistent.  I would be surprised if the further working out of this
interpretation did not change many of its details (for example the shapes
of the spacetime regions which enter), but I also feel that it is broadly
on the right track.


\bigskip
\noindent{\bf Criteria for Preclusion (unconditional case)}
\Nobreak

Let us assume, for the moment, a non-dynamical spacetime $M$ with a fixed
causal structure, and consider an event $E$ {\it defined in} a region
$R\subseteq M$.  Here ``an event'' means just a set of histories
($E\subseteq \H$), and describing it as ``defined in $R$'' will mean that
the properties which determine whether a given history $\gamma$ belongs to
$E$ or not pertain only to the portion of $\gamma$ within $R$ (specifically
to $\gamma$'s {\it intersection} with $R$ if it is a set of world lines,
its {\it restriction} to $R$ if it is a field, etc).  We now ask, ``When is
such an event $E$ precluded?''

An answer given along the lines of [16] would be something like,
``$E$ is precluded when $|E|=0$ (or $<\epsilon$) and we {\it can}
operationally distinguish $E$ from its complement, but we {\it cannot}
operationally distinguish subsets of $E$ from each other''.  (For example,
in a two-slit experiment, the arrival of the electron at a dark band on
the screen would be precluded because it has measure zero, and we {\it can}
tell whether the electron has landed there rather than elsewhere, but we
{\it cannot} tell which of the two slits it traversed.)

The trouble with this criterion, of course, is that it appeals to our
possibility of knowledge, which is at once subjective and
vague.\footnote{*}
 {which is not to say that the criterion could not still be perfectly
  satisfactory in many situations.}
On the other hand, it won't do just to drop all qualification and say that
$E$ is precluded whenever its measure is sufficiently small.  That
criterion would be objective and well-defined, but it would in general lead
to the absurd result that {\it everything} would be precluded.

A striking illustration of this difficulty is furnished by the three-slit
experiment referred to earlier, which we may idealize for present purposes
as a source emitting spinless particles which impinge on a diffraction
grating with three slits.  
(See figure 2.)
Let $P$ be a spacetime
region---idealized as a point---which is aligned with the central slit and
consequently sits within a ``bright band'' of the diffraction pattern.  For
such a point, the rule (1) yields a nonzero value for the measure
$|\{1,2,3\}|$ of the set of all world lines that arrive at $P$ via any one
of the slits, and we know, in fact, that if we look for the particle at $P$
by placing a detector there, we will often find it.  On the other hand, we
can choose the separation between the slits so that, when taken in pairs
$\{1,2\}$ or $\{2,3\}$, the amplitudes cancel, and correspondingly, the
measures of $E=\{1,2\}$ and of $F=\{2,3\}$ will vanish: $|E| = |F| = 0$.
An unrestricted preclusion rule would then entail that the actual history
could belong neither to $E$ nor to $F$, whence it could not arrive at $P$
at all---a false prediction.  More generally, one can typically embed any
given history $\gamma$ in a subset $S \subseteq \H$ of zero measure, whence
every possibility without exception would be ruled out by an unrestricted
preclusion rule.\footnote{$^\dagger$}
{In a certain sense the same problem is present even classically (where for
example, each single Brownian path taken individually is of measure zero),
but the contradiction in the quantum case is made worse by the effects of
interference.  If, despite this, one could somehow ``learn to live with the
contradiction'', as one does in the case of classical probability theory,
then the remainder of this paper would be superfluous.}

\epsfysize=2.0in
 \centerline{\vbox{\epsfbox{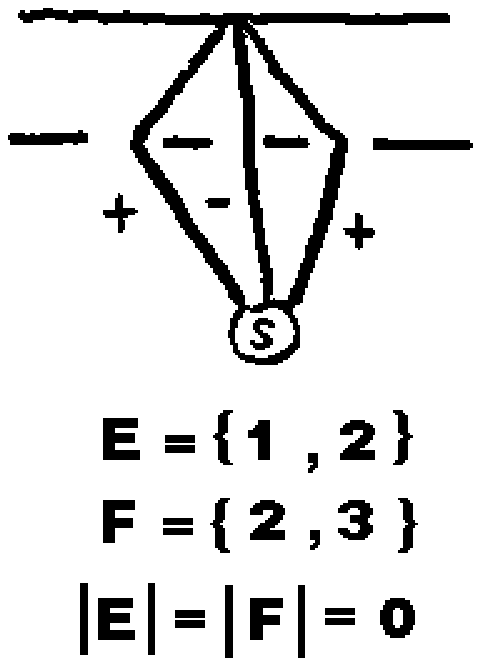}}}
\centerline{{\bf Figure 2} Possible histories in three-slit diffraction}
\vskip .25in

To avoid such nonsensical conclusions, we are obliged to place further
limits on the application of the preclusion concept.  In seeking
inspiration for such a modified criterion, it is natural to refer mentally
to measurement situations; for they are one arena where contradictions of
the above sort are avoided.  We may therefore suspect that there is
something special about measurement situations which, if correctly
identified, would provide a prototype on which a more general, objective
interpretation of the quantum measure could be modeled.  But what is this
special feature?  If your answer turns on ``amplification'' and the
decohering of macroscopically distinct alternatives, you will be led in the
direction of a ``consistent histories'' interpretation of quantum
mechanics.  If, on the contrary, you fasten on the correlation which a
measurement sets up between observer and observed, you will be led to the
sort of scheme I am about to propose.\footnote{*}
 {This may be a good place to comment on the relation of my interpretation
 to the consistent histories scheme, where by the latter I mean especially
 the version described in [7].  The two approaches agree on the
 spacetime character of reality; both take the measure $|\cdot|$ to be the
 fundamental quantity governing the dynamics; and both attempt to honor the
 ``principle of realism-objectivity'' by working with ``beables''
 (properties of histories) rather than ``observables''.  As presented in
 [7] and elsewhere, the realistic aspect is not so much stressed,
 but this seems more a matter of philosophical predisposition than of a
 genuine difference in the schemes.  The real difference lies in the
 interpretation of the measure.  The ``consistent historians'' would in
 effect reduce the quantum sum-rule $I_3=0$ back to the classical one
 $I_2=0$ by limiting the domain of $|\cdot|$ to a suitable family of
 subsets (``coarse-grained histories'').  In this way ``consistency'' is
 achieved, in the sense that one has returned to classical probability
 theory, but direct contact with the micro-world tends to be lost (because
 decoherence tends to require macroscopic objects), and the principle of
 the single world is sacrificed to the extent that the choice of
 coarse-graining is non-unique, leading either to conflicting probability
 predictions or the admission of several distinct ``worlds'', each with its
 own dynamical laws [17].  In contrast, my approach
 embraces the non-classical behavior of the measure as that which makes
 quantum mechanics what it is, and tries to interpret $|\cdot|$ directly
 via a preclusion rule which is strong enough to justify standard
 applications of the quantum formalism, but weak enough to avoid the kind
 of contradiction exposed above.  Despite these differences there are many
 parallels and connections between the schemes, some of which should be
 apparent in the following pages.}

This scheme, specifically, will be founded on the idea that the quantum
measure acquires predictive content only when variables belonging to two
kinematically independent subsystems become correlated, or more accurately,
since neither ``kinematic independence'' nor ``subsystem'' has any definite
meaning in general, when there occur {\it correlations between events
which} (in the technical sense introduced above) {\it are defined in
spacelike separated regions of} $M$.

This is the basic idea, but its most straightforward implementation does
not entirely eliminate the problem of ``overlapping preclusions'', of which
the 3-slit diffraction described above is only the simplest instance.  More
complicated instances remain, where the problems arise from specious
correlations between non-interacting systems containing ``null events''
(events of measure 0), or from correlations of the sort that occur in the
Kochen-Specker version of the ``EPRB''-experiment.  There is no space here
to explain these problems in detail (a fuller account will appear
elsewhere), but the attempt to frame a preclusion criterion which can
exclude both types of difficulty leads one to introduce a third spacetime
region, whose effect is not only to remove the unwanted preclusions, but
also (as an unanticipated benefit) to broaden the range of inferences one
can draw about the past 
in situations where
the new criterion {\it is} satisfied.  (Incidentally, most of the problem
situations just alluded to have implications for consistent histories as
well, where they typically provide new types of examples of mutually
inconsistent coarse-grainings.  For instance, in the 3-slit situation
above, $\{1,2\} \sqcup \{3\}$ and $\{1\}\sqcup \{2,3\}$ are both
``consistent'' coarse-grainings in themselves, but for the first, only
alternative $\{3\}$ can happen, whereas for the second, only alternative
$\{1\}$ can happen.)

The specific criterion or {\it preclusion rule} which emerges may be stated
formally as follows 
(see figure 3).  
Let $I$, $II$ and $III$ be a
triple or ``triad'' of spacetime regions, and for each $R= I, II, III$ let
$E_i^R$ ($i=1\ldots n$) be a system of $n$ events defined in $R$.  Let
regions $I$ and $II$ be spacelike to each other, with $III$ equal to the
common future of $I \cup II$; or in formulas:
$$
        I \; \natural \; II   
           \qquad\qquad{\rm and}\qquad\qquad
        III = {\rm future} (I) \cap {\rm future} (II),
$$
where `$R \,\natural\, S$' means that every element of $R$ is spacelike to
every element of $S$, and `future $(R)$' denotes the set of all points
which are to the future of {\it every} element of $R$.  For each
$R=I,II,III$, let the $E_i^R$ partition ${\cal H}$ (meaning that $\H =
\sqcup_i E_i^R$ is their disjoint union; such systems of events are often
called ``exclusive and exhaustive'').  Further, let ${\~{E}}_i^R$ denote an
arbitrary ``sub-event'' (i.e. subset) of $E_i^R$, also defined in $R$ .
Call a triple $i$, $j$, $k$ ``diagonal'' when $i = j = k$, and
``off-diagonal'' otherwise.  Then, the criterion comprises two assertions.
First, if
$$
    | {\~ E}_i^I \cap  {\~ E}_j^{II}  \cap  {\~ E}_k^{III} | = 0
   \eqno(5)
$$
in all off-diagonal cases and for all choices of sub-events
${\~{E}}_i^R$, 
{\it then every off-diagonal event $E_i^I \cap E_j^{II} \cap E_k^{III}$ is
precluded}.  Second, if in addition to (5) we have
$$  |E_i^I \cap E_i^{II} \cap E_i^{III}| = 0 \eqno (6)$$
for some particular $i$, {\it then that $E_i$ itself is precluded}.
(Notice that in the presence of (5), (6) entails 
$E_i^I = E_i^{II} = E_i^{III} = 0$ by the axiom (3).)  Moreover,
these implications will, if necessary\footnote{$^\dagger$}
 {That such a ``fuzzing'' might conceivably be dispensed with, is due to the
 fact that the measure of a nontrivial quantum alternative can vanish
 exactly, unlike in classical probability theory.},
be taken to hold when ``$= 0$'' is replaced by ``$< \epsilon$, for
sufficiently small $\epsilon$''.

\epsfysize=2.0in
 \centerline{\vbox{\epsfbox{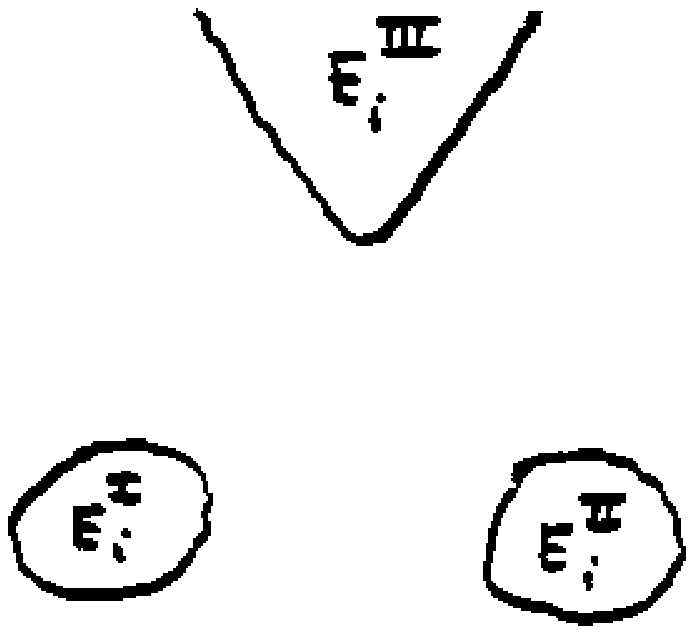}}}
\par\nobreak
\centerline{{\bf Figure 3} A triad of spacetime regions}

\vskip .25in

The effect of this rule is to delineate a class of situations in which one
can uphold the idea that sets of zero measure are precluded, even if (as we
have seen) this idea cannot consistently be maintained in general.
Specifically, the criterion tells us that certain three-way correlations
which are ``present in the measure'' will in fact be present in the history
itself\footnote{*}
 {The indirect wording of the criterion, which expresses a correlation as
 the negation of its negation, could be avoided if the opposite of
 preclusion could be defined directly, but this seems to be impossible
 because there is no analog of being of probability 1 for the quantum
 measure (which is unbounded above, due to the possibility of {\it
 constructive} interference).},
and that when such a correlation is in force, not only are all the
``off diagonal'' possibilities precluded, but also any of the ``diagonal''
ones which are themselves of zero measure.

\vbox{
 \epsfysize=1.8in
 \centerline{\vbox{\epsfbox{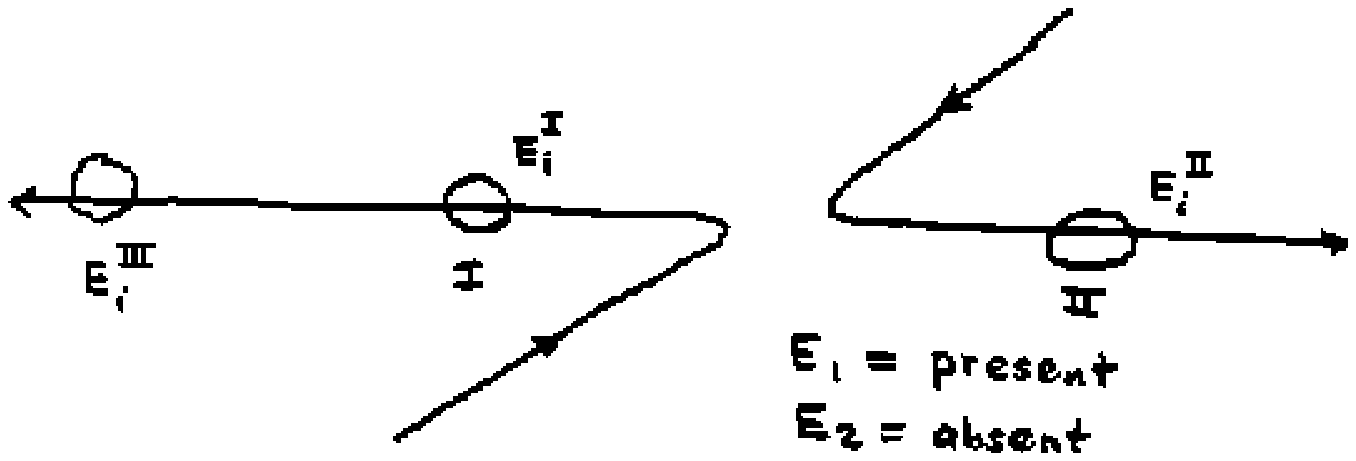}}}
\centerline{{\bf Figure 4 a} Correlations in electron-electron scattering }
}
\vskip .25in
We can illustrate the meaning of our preclusion criterion with an idealized
account of the scattering of low-energy polarized electrons in their center
of mass frame 
(see Figure 4).  
In that situation the amplitude for
the electrons to emerge other than in approximately opposite directions is
zero, as is the amplitude for them to scatter at 90 degrees (due to the
fermionic cancelation between ``exchanged'' trajectories).  Then, let us
choose our regions $I$ and $II$ to be situated simultaneously in time, and
spatially opposite each other with respect to the scattering center; and
let their extension be large enough so that the corresponding ``uncertainty
principle disturbances'' of the electrons' energy and momentum are
negligible.  Let us take event $E_1^I$ to be the {\it presence} of an
electron in $I$, and $E_2^I$ to be the complementary possibility (absence
of any electron there), with the $E_i^{II}$ defined analogously.  For the
corresponding pair of events in region $III$, we can take the presence or
absence of an electron at some convenient location within $III$ to which
the continuation of a scattering trajectory through region $I$ (say) would
lead.

 \epsfysize=2.0in
 \centerline{\vbox{\epsfbox{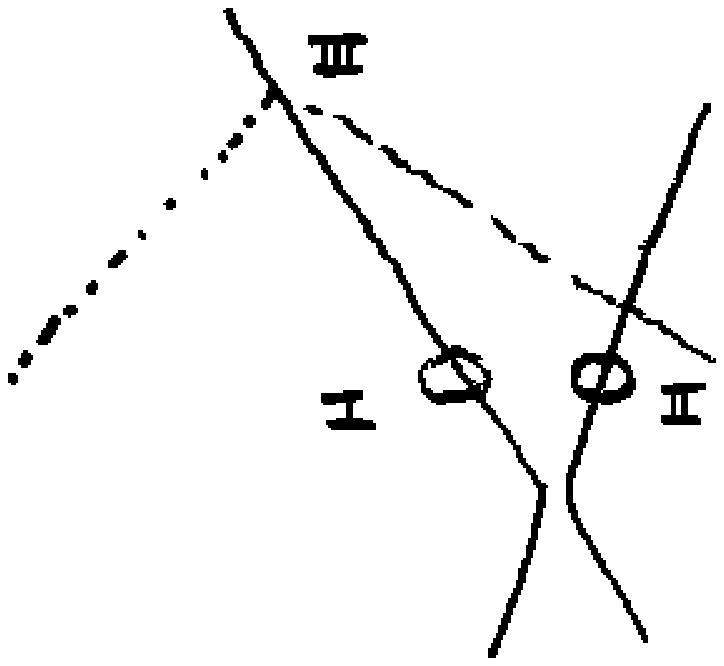}}}
\par\nobreak 
\centerline{{\bf Figure 4 b} The same scattering as a spacetime diagram}
 \vskip .25in

  Then the first clause of the preclusion criterion will be satisfied,
whence we can assert that either there will be electrons in both regions or
in neither (and the former happens if and only if the electron traversing
region $I$ also is found at the ``confirming'' location within region
$III$).  By appealing to a sufficient number of such triads, we can then
claim that the electrons do in reality emerge opposite each other (to an
accuracy governed by the region size).  Recall that this assertion concerns
the trajectories themselves, and is to be regarded as meaningful whether or
not the regions are provided with electron detectors.\footnote{$^\dagger$}
{A slightly more elaborate argument along similar lines implies that, in a
 diffaction experiment, the electron actually avoids the ``dark'' regions,
 whether or not a detecting screen is present.  Such a conclusion might
 seem even more striking than the one just discussed for electron
 scattering.  However it also turns out to be less ``stable'' against small
 changes to the formulation of the preclusion criterion.}

For an application of the second clause of the preclusion criterion, we can
now choose regions $I$ and $II$ to be at $90^\circ$ and $270^\circ$
respectively, and then we predict that no electron will appear in either
region.  Once again, this is a prediction about the electrons themselves,
independently of any apparatus which we might install to help confirm the
prediction experimentally.

(The illustration of the preclusion criterion we have just given deals
directly with the behavior of microscopic objects.  For macroscopic objects
such as measuring instruments, the possible triads are much more numerous
and harder to analyze completely, but it appears plausible that every
macroscopic event is enmeshed in a wealth of triads in such a manner that
one can conclude that, in an actual ensemble of repeated measurements, it
is precluded to obtain a set of results very far from those predicted by
application of the usual probability rules.  Here one would in effect be
using the concept of preclusion in its classical form, relying on the
independence of the separate trials and the decoherence of the individual
outcomes to make the law of large numbers work as it does classically.)

Although there is no room for a thorough motivation of the above preclusion
rule (or ``criterion''), a few comments and observations seem in order.

First, the criterion is time-asymmetric due to the placement of region
$III$ to the future, rather than the past, of the other two regions.
Perhaps this asymmetry corresponds in some manner to the $T$-asymmetry of
``wave-function collapse'' in traditional interpretations, or perhaps it is
more fundamental, since the collapse-asymmetry is well-known to disappear
when one expresses the probabilities directly in terms of products of the
corresponding projections operators [18].

Second, the correlations  which figure in our criterion are obvious
generalizations of correlations  which routinely occur in the course of
ordinary 
measurements, where $E_i^I$ (say) would be a particular property of the
``micro-object'', $E_i^{II}$ the corresponding response of the
``instrument'', and $E_i^{III}$, for example, a record kept of the result.
The only aspect of our rule that could not be directly motivated by this
paradigm is the requirement that regions $I$ and $II$ be spacelike
separated.  (Notice however that the asymmetry in ordinary measurements
between observer and observed is {\it not} mirrored by any asymmetry
between regions $I$ and $II$ in our preclusion rule.)

But why did we need region $III$ at all?  As mentioned above, the
motivating example is a Gedankenexperiment in which 117 Stern-Gerlach
analyzers for each of two correlated spin-1 particles are set in place with
each analyzer followed by a ``re-combiner'' which undoes its effect.  With
appropriate settings for the analyzers one predicts more overlapping
correlations than the particles can consistently satisfy.\footnote{*}
 {The logical contradiction involved here is the same one which shows that
 local hidden variable theories can never reproduce the pattern of perfect
 correlations predicted by quantum mechanics for a pair of suitably
 ``entangled'' spin-1 particles.  This is a stronger contradiction than
 provided by Bell's inequality, because it is not merely probabilistic, but
 operates instead at the level of logic.  That the analysis of [19] can
 be used to strengthen Bell's result in this way was pointed out in
 [20], and made the basis of a Gedankenexperiment utilizing
 Stern-Gerlach spin-analyzers and recombiners in [21] (see also
 [22]).  To exhibit the contradiction experimentally, one must set the
 analyzers to 117 different spin-axes during the course of the experiment.
 An analogous Gedankenexperiment involving three spin-1/2 particles but
 only two distinct settings for each analyzer was given subsequently in
 [23].}
The effect of requiring correlation with
region $III$ as well, is to force a given $I-II$ correlation to ``persist''
long enough that some of the others with which it would conflict can no
longer have {\it their} regions $I$ and $II$ mutually spacelike.

Region $III$ is also helpful when we want to draw conclusions about the
past in the manner of geology.  Since none of the exploits of the
dinosaurs were spacelike to our discovery of their fossils, it wouldn't
work to take these events as belonging to regions $I$ and $II$, but we {\it
can} associate the fossils to region $III$, once it has been introduced
into the scheme.

\bigskip
\noindent{\bf Conditional Preclusion}
\Nobreak

Both in science and in daily life, the predictions we make rest on
presuppositions about the past, presuppositions which in practice derive
partly from knowledge obtained through observation and partly from the more
or less definite assumptions we make about initial conditions (which we
might imagine as being at the time of the big-bang or the immediately
preceding quantum era).  If all valid assertions of preclusion could be
founded on the initial conditions alone, then there would be no need for a
separate criterion for conditional preclusion, since the latter would just
be a special case of absolute preclusion.  (Conditional preclusion of an
event $A$ given the (earlier) event $B$, would just signify absolute
preclusion of their ``conjunction'' $A\cap{B}$.)  However, there seems
little prospect in practice of reducing the necessary input to cosmological
initial conditions, and I suspect it could not even be done in principle.
If so, we need a logically separate rule for conditional preclusion, and I
offer here a very preliminary suggestion of how it should go.

No doubt it would be adequate, from a practical point of view, to employ
the rule from classical probability theory that we incorporate our
knowledge of the past just by {\it relativizing} the measure we use for
future events to that knowledge; that is, in computing the measure $|S|$ of
a future event we would restrict the sum in equation (1) to histories
$\gamma$ compatible with our knowledge.  However satisfactory such a rule
might be for practical purposes, though, it does not appear possible to
restate it objectively without coming into conflict with the principle of
the single world.  Take two-slit
interference, for example.  We can certainly compute that arrival at a
detector in a dark band is precluded if we apply our criterion at a moment
just after the electron leaves the source, but what about just after it has
passed through one of the slits?  At that stage the interfering alternative
is no longer available, but by definition the preclusion cannot have gone
away, even though the relativized measure of the event ``arrival at the
detector'' is no longer zero.\footnote{*}
 {This is doubtless a major reason why people have felt driven to abandon
 realism in interpreting quantum mechanics.}
In fact, if we were to specify the electron's world-line with full
precision, and then relativize to the corresponding subset of histories, it
would be as if we had forced the electron ``by thought alone'' into a
position eigenstate, and our ensuing predictions would be completely
erroneous.

To state the difficulty more generally, it seems that we should be able to
make statements like: ``If the actual history $\gamma$ is an extension of
the (particular) initial segment $\hat\gamma$, then such and such a set $S$
of possible 
future developments of $\hat\gamma$ is precluded''; yet adopting as the
criterion for such statements our earlier preclusion rule with the measure
restricted to those histories sharing the common initial segment
$\hat\gamma$ yields the wrong results.

What seems to be needed is a way to bring some of the ``bypassed''
alternatives back into the picture, or in other language, to ``fuzz out''
the initial segment $\hat\gamma$ on which we condition the definition of
the relativized measure.  It appears reasonable to take this fuzzing to be
induced by a coarse-graining of the gravitational field\footnote{$^\dagger$}
 {In causal set theory such coarse graining amounts to passing to a
 randomly chosen subset of the true causal set [24]},
since that plausibly would have the desired effect, in the two-slit case
for example, of allowing a wide latitude to the electron trajectory while
still keeping the diffraction grating well-localized.  If fuzzing via
gravitational coarse-graining is the right approach, there still remains
the question of how much fuzzing to perform, to which a reasonable answer
might be that preclusions arrived at by {\it any} fixed degree of fuzzing
are valid.

\bigskip
\noindent{\bf When Spacetime is Dynamical...}
\par\nobreak

When spacetime is dynamical (i.e. in quantum gravity) our preclusion rule
becomes meaningless unless there is a way to specify in advance the triads
of spacetime regions in terms of which it is phrased.  Indeed, one might
think to discern an unbreakable vicious circle here, because it seems
impossible to locate a region not yet in being without having the kind of
advanced knowledge of the geometry which could only be provided via
preclusion statements, whose meaning relies on the ability to locate
spacetime regions in advance.  This circle is related to what is sometimes
called ``the problem of time'', and one might try to resolve it by finding
a way to identify the regions in question by means of invariant properties
like curvature-invariants or their discrete analogs.  Such an approach
might or might not appear promising, but here I only want to point out,
without trying to provide a definitive answer to the question that, {\it in
connection with conditional preclusion}, there exists another, much more
specific way to ``locate'' future regions, by ``projecting forward from
what already exists''. 

 \epsfysize=2.0in
 \centerline{\vbox{\epsfbox{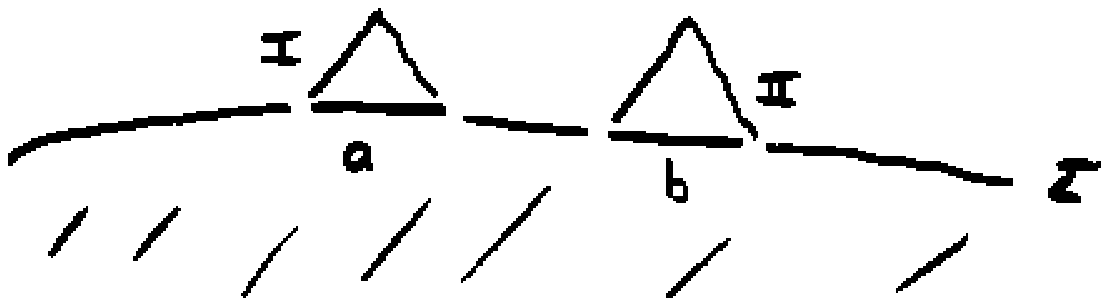}}}
\centerline{{\bf Figure 5} Locating future regions in a dynamical spacetime}
 \vskip .25in

As an illustration, let $M$ (the history) be a Lorentzian manifold, and let
us condition on the property that $M$ contain some definite initial segment
with future boundary $\Sigma$.  (See Figure 5.)  Then if $a$ and $b$ are
any two disjoint regions on $\Sigma$, their future domains of dependence
will necessarily be regions of
$M$ which are disjoint and spacelike separated.  Hence they can be used for
regions $I$ and $II$ of a triad, with region $III$ then being defined---as
always---in terms of $I$ and $II$.  In formulas, $I = D^+(a)$, $II=D^+(b)$,
$III = \future(I)\cap\future(II)$.  (Notice that in the causal set case,
exactly the same definitions work if we interpret $D(a)$ as the ``double
causal complement'' of $a$ (cf. [25]), $D(a) = a'' 
= \{x|\forall y, y\natural a \implies x \natural y\}$, 
where $R' := \{x| \forall r\in R, r\natural x \}$.)  
Clearly, many other constructions of a similar type are possible, and would
also suffice to specify ``in advance'' definite regions $I$ and $II$ to the
future of $\Sigma$.

\bigskip
\noindent{\bf What More is Needed?}
\Nobreak

What more is needed in order to give us confidence that the interpretive
scheme outlined above is truly adequate to its task?  Several further steps
seem called for.

First of all one must look for remaining contradictions which even our more
restrictive preclusion criterion may still not have eliminated.  I would
conjecture that none will be found, but on the other hand, I have no good
idea how one might go about providing a proof for this conjecture.  Perhaps
one can be satisfied if enough attempts to construct a contradiction fail,
much as one has gained confidence in the consistency of Zermelo-Fraenkel
set theory as the anticipated contradictions have failed to materialize.  A
related question is whether the introduction of a small number $\epsilon$
into our preclusion criterion is really needed, or whether it suffices to
assert preclusion for sets whose measure vanishes exactly.  Perhaps this
question will prove easier to settle than that of consistency in general.

One should also think through a number of everyday situations (including
measurement situations) to see if enough triads are present to justify the
kinds of conclusions we normally draw, not only in quantum mechanical
situations, but more generally.  For example, can we conclude that the
starlight we see when we look out at the sky was emitted by actual stars
situated along our past light-cone?  Here there do seem to be triads of the
required type as illustrated in Figure 6, where our detection of the light
occurs in region $II$, and region $III$ contains, say, our memory of just
having seen a star.  A peculiarity of the preclusion rule in this situation
is that event $I$ cannot be taken to be the emission of the light itself,
as that would not be spacelike to region $II$.  Instead, we can choose
region $I$ just to the future of the light's path, with the corresponding
event being the presence of the star there.  What such a triad lets us
conclude is that a star was present {\it just after} the earliest photon we
received would have been emitted.  Although the logic here is a bit
different from what one is used to, the conclusion itself is
indistinguishable in practice from what we normally obtain.

\epsfysize=2.0in
 \centerline{\vbox{\epsfbox{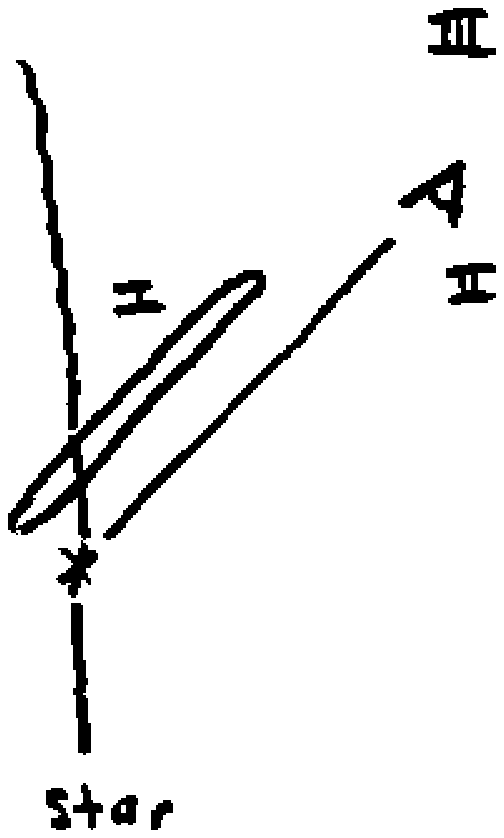}}}
\centerline{{\bf Figure 6} Stargazing}
\vskip .25in

A third desideratum, of course, would be the sharpened conception of
fuzziness needed to render the notion of conditional preclusion precise.
And once this is found, it should be checked whether the resulting
criterion for conditional preclusion is adequate to its task, which
includes a requirement that it be free of the kind of inconsistency we
discussed above (of too many possibilities precluded).

Fourth, some further developments of quantum measure theory, though not
logically necessary for the interpretive framework, would also be
desirable.  In particular, it would be nice to understand what extra
conditions on the measure correspond to unitary evolution, i.e. what
conditions would cause the measure to be expressible in the form (1)
with $|S|$ independent of the choice of $T$.

Finally, if more tasks are needed, there is the one for which all of the
above is just preparation: find the right space of histories and the
appropriate measure on it to describe quantum gravity!

\bigskip\noindent

I would like to express my gratitude to R. Salgado for his help with the
diagrams. 
This research was partly supported by NSF grant PHY-9307570.

\singlespace					   
\vskip 0.5truein
\centerline           {\bf References}
\nobreak
\medskip
\noindent
\parindent=0pt
\parskip=10pt

[1]
 R.D.~Sorkin
 ``Forks in the Road, on the Way to Quantum Gravity'', talk 
   given at the conference entitled ``Directions in General Relativity'',
   held at College Park, Maryland, May, 1993,
   (Syracuse University preprint SU-GP-93-12-2)

[2] 
Weyl, H., ``Wissenschaft als Symbolische Konstruction des Mensch\-en'',
        Er\-anos-Jahrbuch 1948, vol. 16: {\it Der Mensch}
        (Z\"urich, Rhein-Verlag, 1949)

[3]
R.P. Feynman,
 ``Spacetime approach to non-relativistic quantum mechanics'' 
   {\it Reviews of Modern Physics} {\bf 20}:367-387 (1948)

[4]
 R.D.~Sorkin, 
``On the Role of Time in the Sum-over-histories Framework for Gravity'',
    paper presented to the conference on The History of Modern
      Gauge Theories, held Logan, Utah, July, 1987; 
        published in  
          {\it Int. J. Theor. Phys.} {\bf 33}:523-534 (1994)

[5]
Sinha, Sukanya and R.D. Sorkin
 ``A Sum-Over-Histories-Account of an EPR(B) Experiment''
    {\it Found. of Phys. Lett.}, {\bf 4}, 303-335, (1991); and\cut
 R.D.~Sorkin, 
``Problems with Causality in the Sum-over-histories Framework
      for Quantum Mechanics'',
   in A. Ashtekar and J. Stachel (eds.), 
  {\it Conceptual Problems of Quantum Gravity} 
   (Proceedings of the conference of the same name, 
       held Osgood Hill, Mass., May 1988), 217--227 
   (Boston, Birkh\"auser, 1991)

[6]  
R.D.~Sorkin, 
   ``Spacetime and Causal Sets'', 
      in J.C. D'Olivo, E. Nahmad-Achar, M. Rosenbaum, M.P. Ryan, 
               L.F. Urrutia and F. Zertuche (eds.), 
      {\it Relativity and Gravitation:  Classical and Quantum,} 
       (Proceedings of the {\it SILARG VII Conference}, 
       held Cocoyoc, Mexico, December, 1990), pages 150-173,
       (World Scientific, Singapore, 1991)

[7] 
Hartle, J.B., ``The Quantum Mechanics of Cosmology'', in {\it Quantum
 Cosmology and Baby Universes: Proceedings of the 1989 Jerusalem Winter
 School for Theoretical Physics}, eds. S. Coleman et al. (World
 Scientific, Singapore, 1991); for brief reviews of the main ideas see\cut
J.B.~Hartle, "The Spacetime Approach to Quantum Mechanics'', 
in 
{\it Proceedings of the Symposium for Louis Witten, April 4-5, 1992},
or in 
{\it Proceedings of the IVth Summer Meeting on the Quantum Mechanics of
Fundamental Systems}, Centro de Estudiois Cientificos de Santiago,
Santiago, Chile, December 26-30,1991,
or in 
{\it Proceedings of the International Symposium on Quantum Physics and the
Universe}, Waseda University, Tokyo, Japan, August 23-27, 1992;
and \cut
J.B.~Hartle, ``The Quantum Mechanics of Closed Systems'', 
in 
{\it Directions in General Relativity, Volume 1: A Symposium and Collection
of Essays in honor of Professor Charles W. Misner's 60th Birthday}, ed. by
B.-L.~Hu, M.P.~Ryan, and C.V.~Vishveshwara, Cambridge University Press,
Cambridge (1993).
%

[8]
Caves, C.M.,
``Quantum mechanics of measurements distributed in time. A path-integral
  formulation'', {\it Phys. Rev. D} {\bf 33}, 1643 (1986)

[9] See for example 
   E.B.~Dynkin, 
  {\it Markov Processes}
  (Academic Press, 1965)

[10]
W. Heisenberg,
 {\it The Physical Principles of the Quantum Theory}
  (translated by C. Eckart and F.C. Hoyt)
   (Dover 1930)

[11]
 R.D.~Sorkin,
 ``Quantum Mechanics as Quantum Measure Theory'',
   {Mod. Phys. Lett.} {\bf A9}, No. 33 (1994), p. 3119
   (available from the gr-qc bulletin board as Paper: gr-qc/9401003)

[12] 
D.~Meyer, 
``Quantum computation and beyond'', 
 Colloquium at Syracuse University, 
 delivered 1 November 1994.

[13] 
A.O.~Caldeira and A.J.~Leggett,
{\it Physica A}{\bf 121}:587 (1983) and {\bf 130}:374(E) (1985)  

[14]
B.L.~Hu, J.P.~Paz and Y.~Zhang,
  ``Quantum Brownian motion in a general environment. II. Nonlinear
    coupling and perturbative approach'',
    {\it Phys. Rev. D} {\bf 47}:1576-1594 (1993)

[15] 
R. Geroch, 
``The Everett interpretation", 
{\it No\^{u}s} {\bf 18}:617 (1984)

[16] 
R.P.~Feynman, R.B.~Leighton and M.~Sands,
  {\it The Feynman Lectures on Physics, vol. III: Quantum Mechanics}
  (Addison--Wesley, 1965)

[17]
M.~Gell-Mann and J.B.~Hartle,
  ``Equivalent Sets of Histories and Multiple Quasiclassical Domains'',
    preprint UCSBTH-94-09


[18]
Y.~Aharonov, P.G.~Bergmann and J.L.~Lebowitz,
  ``Time Symmetry in the Quantum Process of Measurement'',
    {\it Phys. Rev.} {\bf 134}:B1410-B1416 (1964)

[19]
Kochen, Simon and E.P. Specker,
``The Problem of Hidden Variables in Quantum Mechanics'',
  {\it Journal of Mathematics and Mechanics} {\bf 17}:59-87 (1967)

[20]
A. Stairs,
{\it Phil. Sci.} {\bf 50}:578 (1983); and private communication

[21]
R.D.~Sorkin, 
 ``The Kochen-Specker Experiment'',
 unpublished Colloquium address at the University of Wisconsin, Milwaukee, 
 delivered May 12, 1986.

[22]
See the first paper in reference [5] above, 
footnote 4 on page 333 therein.

[23]
 D.M.~Greenberger, M.~Horne and A.~Zeilinger, in
{\it Bell's Theorem, Quantum Theory and Conceptions of the Universe},
ed. M.~Kafatos,
(Kluwer, Dordrecht, 1989)

[24]    
Bombelli, L., J.~Lee, D.~Meyer and R.D.~Sorkin, 
 ``Spacetime as a causal set'', 
    {\it Phys. Rev. Lett.} {\bf 59}, 521 (1987)

[25]  
R. Haag,
 {\it Local Quantum Physics: Fields, Particles, Algebras},
	(Springer-Verlag, 1992), see section III.4.1.

\end